\documentclass[a4paper,11pt]{article}
\usepackage{pos}
\usepackage{subfigure}
\usepackage{siunitx}
\usepackage{comment}

\title{Searches for baryon number violation \\ in the HIBEAM-NNBAR experiment \\ at the European Spallation Source}

\author*[a,b]{Bernhard Meirose}

\affiliation[a]{Department of Physics, Stockholm University, \\
106 91 Stockholm, Sweden}

\affiliation[b]{Department of Physics, Lund University,\\
P.O Box 118, 221 00 Lund, Sweden,}

\emailAdd{bernhard.meirose@fysik.su.se}
\emailAdd{bernhard.meirose@hep.lu.se}

\abstract{The HIBEAM-NNBAR program is a proposed two-stage experiment at the European Spallation Source (ESS) designed to search for baryon number violation, which is – together with C and CP violation – one of the three fundamental Sakharov conditions to explain the observed baryon asymmetry of the Universe. Taking advantage of the ESS' unique capabilities as the future brightest neutron source, the experiment will make high sensitivity searches for neutrons converting into antineutrons and/or sterile neutrons.}

\FullConference{%
  21st Conference on Flavor Physics and CP Violation (FPCP 2023),\\
  May 29 - June2, 2023\\
  IP2I - Lyon University, Lyon, France
}

\begin{document}
\maketitle

\section{Introduction}

The HIBEAM-NNBAR program \cite{Addazi:2020nlz,Abele:2022iml} is a multi-stage series of experiments to be conducted at the European Spallation Source (ESS). The primary aim is to investigate the possibility of neutrons undergoing conversion or oscillation into antineutrons ($n\rightarrow\bar{n}$), as well as the potential transformation of neutrons into sterile neutrons ($n\rightarrow n'$). The detection of such phenomena would imply baryon number violation (BNV), a crucial criteria outlined by Sakharov for baryogenesis ~\cite{Sakharov:1967dj}. Neutron transformations represent a promising avenue for testing the validity of the Standard Model (SM) of particle physics beyond the neutrino sector.

\section{Motivation}
\subsection{Why baryon number violation?}

Baryon number ($\mathcal{B}$) conservation was first postulated by Hermann Weyl in 1929, primarily as a way to explain the apparent stability of matter through proton stability, similarly to how charge conservation is used to explain electron stability \cite{Weyl:1929fm}.

Although this explanation might appear appealing, the Standard Model (SM) itself does not require any such quantity to be conserved \textit{a priori} and there is no symmetry in the SM requiring baryon number conservation. As it is well known, baryon number is only an accidental symmetry of the SM, meaning that by taking its particle content, the full set of possible renormalizable interactions of the model preserves baryon number, without having it required in the first place. This fact has naturally raised questions about the stability of protons. In contrast to the well-established stability of electrons, which is firmly grounded in the conservation of electric charge (since electrons are the lightest particles with electric charge), the stability of protons lacks an equivalent underlying "fundamental" symmetry. Unlike the electromagnetic gauge invariance that gives rise to electric charge conservation - a genuine local symmetry accompanied by the photon as a gauge boson - proton stability relies on baryon number, which is merely a global symmetry without an associated mediator.

On the other hand, taken together with lepton number ($\mathcal{L}$), the quantity $\mathcal{B}-\mathcal{L}$ is exactly conserved in the SM, and any violation of it would immediately signal the presence of beyond Standard Model (BSM) physics. 

Moreover, baryon number violation in the SM, although technically allowed, can only occur in the non-perturbative regime of the electroweak theory and is only expected to occur at exceptionally high temperatures, being thus unobservable today. Indeed, BNV has long been considered \cite{Sakharov:1967dj} one of the fundamental conditions to explain the observed baryon asymmetry of the Universe (BAU) and might be the key link between dark matter and ordinary baryonic matter, e.g, if baryon number is apparently violated, but conserved across dark and visible sectors \cite{Davoudiasl:2012uw, Gardner:2022yeq}.

The observation of BNV in laboratory experiments would, therefore, not only be a solid and unmistakable signal of BSM physics, but a breakthrough in our understanding of a nonzero cosmic baryon asymmetry. 

\subsection{Why neutron oscillations?}

As experimental limits on proton decay are highly stringent \cite{ParticleDataGroup:2022pth}, it is tempting to think that
these constraints would apply to neutron-oscillations as well. However, proton decay is a $\Delta \mathcal{B} = 1$ process, while neutron oscillations are $\Delta \mathcal{B} = 2$ processes, and as such, can be of complete different physical origin \cite{Davidson:1978pm,Rao:1983sd,Mohapatra:1980qe,Nussinov:2001rb,Arnold:2012sd,Dev:2015uca,Gardner:2018azu,Heeck:2019kgr,Girmohanta:2019fsx,Girmohanta:2020eav} and are also much less constrained \cite{Gardner:2018azu,Heeck:2019kgr}. Crucially, processes that violate baryon number by two units do not need to lead to nucleon decay at all. If one invokes a new scale at which $\mathcal{B}-\mathcal{L}$ is broken, implying new BSM dynamics at play, $\Delta \mathcal{B} = 2$ processes can be generated while leading to small or even vanishing impact on proton decay (see \cite{Berryman:2022zic} and references therein).

Furthermore, even if one disregards proton decay limits, neutron oscillations are still a better/cleaner process to study BNV, since in such processes, \textit{only} baryon number is violated, while single nucleon two-body decays such as $p\rightarrow \pi^0 e^+$ or $p \rightarrow \pi^+ \nu$ always require lepton number violation as well in order to ensure angular momentum conservation. 

Observation of neutron oscillations is therefore the experimental smoking gun of BNV, which is, in turn, a fundamental condition to explain the observed BAU. 

The existence of neutron-antineutron oscillations is also of great importance from a pure fundamental particle physics perspective, as it would demonstrate that the neutron has a Majorana mass, since the existence of fundamental Majorana dynamics is signalled by the breaking of $\mathcal{B}-\mathcal{L}$ symmetry by the baryon number violating process by two units. This is the exact analogous of demonstrating the Majorana nature of the neutrino should neutrinoless double beta decay be observed.

\section{The European Spallation Source}
The European Spallation Source (ESS), currently being constructed in Lund~\cite{Peggs:2013sgv}, will emerge as the world's leading facility for neutron-based research. It will surpass all existing nuclear reactor facilities in terms of neutron flux, offering neutron beams that are up to two orders of magnitude brighter than any other neutron source. ESS is structured as a European Research Infrastructure Consortium (ERIC) and presently comprises 13 member states: Czech Republic, Denmark, Estonia, France, Germany, Hungary, Italy, Norway, Poland, Spain, Sweden, Switzerland, and the United Kingdom. 

The high neutron flux at ESS can be attributed to the fact that it will host the world's most powerful particle accelerator measured in terms of megawatts (MW) of beam on target. It will feature a 62.5 mA proton beam accelerated to 2 GeV. With a pulse structure of 14 Hz, each pulse lasting 2.86 ms, ESS will generate an average power of 5 MW and a peak power of 125 MW. For experiments focused on fundamental physics that prioritize total integrated neutron flux, the ESS concept offers a singular opportunity.

\begin{figure}[tb]
  \setlength{\unitlength}{1mm}
  \begin{center}
  \includegraphics[width=0.60\linewidth, angle=0]{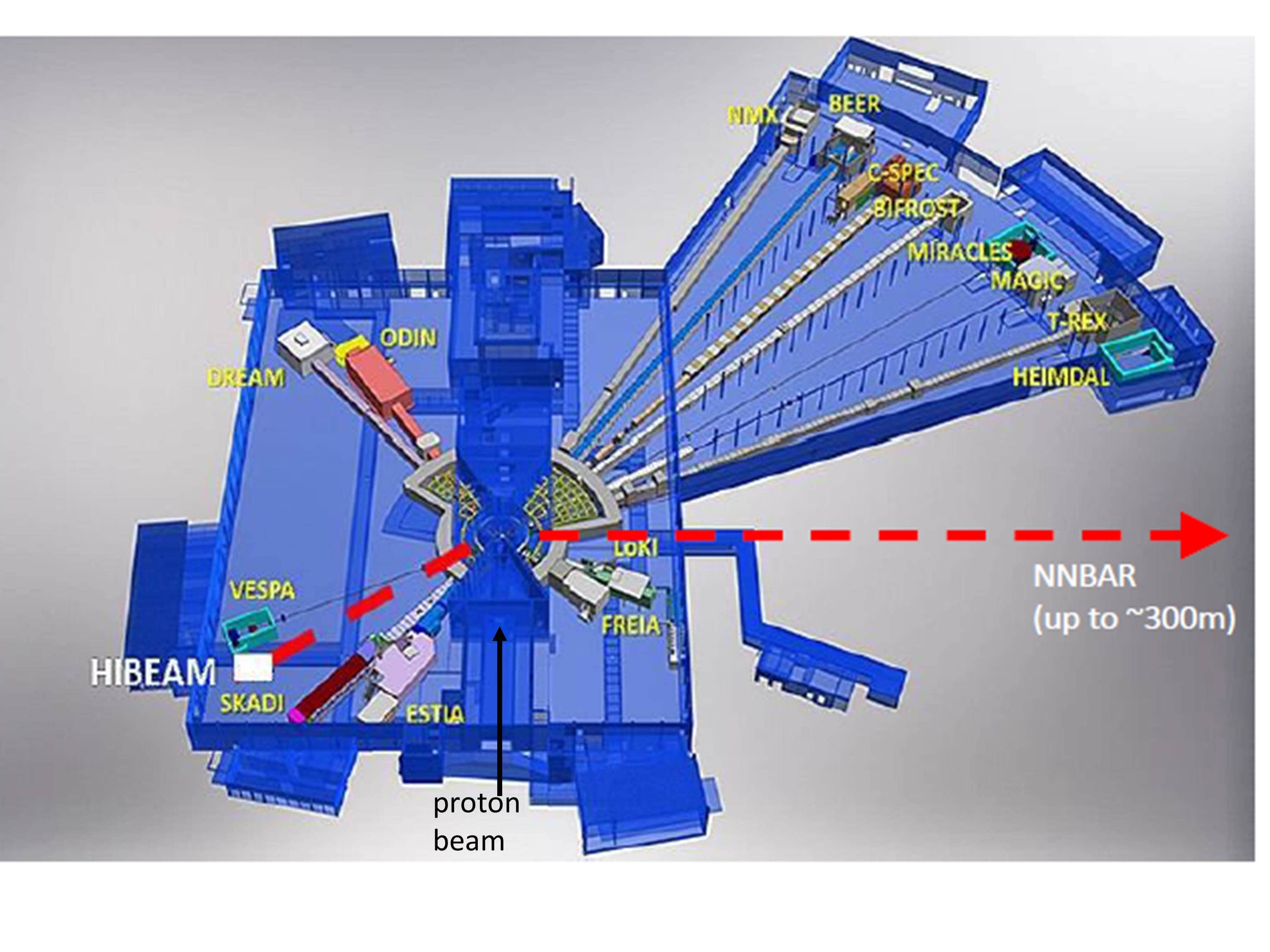}
  \end{center}
  \vspace*{-10mm}  %
  \caption{\footnotesize Overview of the ESS, beamlines and instruments. The locations for the proposed HIBEAM and NNBAR experiments are also shown. }
  \label{fig:ESSinstruments}
\end{figure}

Figure \ref{fig:ESSinstruments} provides an overview of the ESS beamlines and instruments. Currently, there are 15 instruments being constructed at ESS, which represents a subset of the full suite of 22 instruments required for the facility to achieve its scientific objectives, as outlined in the ESS statutes. The anticipated location of the proposed ESS fundamental physics beamline, which would host~HIBEAM, is indicated at beamport E5 in the figure. The prospective beamline leading to NNBAR is also depicted.

\section{The HIBEAM-NNBAR Experiment}

The initial stage of the HIBEAM-NNBAR program is referred to as High Intensity Baryon Extraction and Measurement (HIBEAM) and focuses on investigating processes involving sterile neutrons, induced by non-zero magnetic field values. The assumption is that sterile neutrons exist in a hidden sector and would be impacted by a sterile magnetic field. Depending on the strength of the magnetic field being used, sensitivity increases of more than an order of magnitude can be achieved \cite{Addazi:2020nlz}. Beyond introducing a new opportunity to explore sterile neutron phenomena, a pilot experiment on the conversion between neutrons and antineutrons is also in the scope of the HIBEAM program. This search can potentially match the precision of the previous search at the Institut Langevin (ILL) ~\cite{Baldo-Ceolin:1989vpk}, and even exceed it for an appropriate choice of material in the guide wall \cite{Kerbikov:2018mct,Nes:2019PRL,Nes:2019PLB,Pro:2020PRD,Gudkov:2021wvn}. The second stage (NNBAR), will exploit the ESS Large Beam Port (LBP), a unique component of the ESS facility to search for free neutron-antineutron oscillations, with which the search sensitivity would be increased by three orders of magnitude compared to the ILL search.

\subsection{Free Neutron Oscillation Experiments: HIBEAM and~NNBAR}
\label{review}
\begin{figure}
	\centering
	\includegraphics*[width=0.63\textwidth]{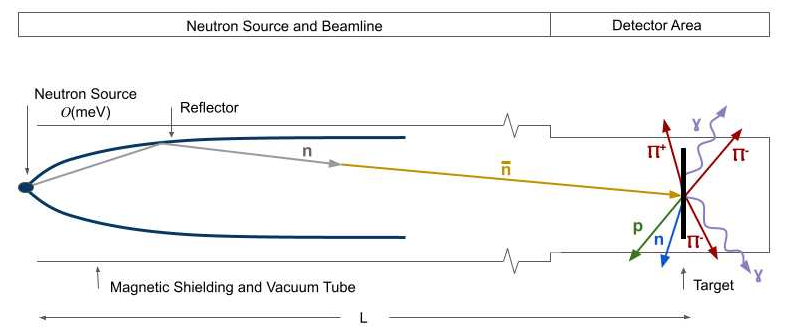}
        \vspace*{-1mm}  %
	\caption{A schematic view of a free $n\rightarrow\bar{n}$ experiment showing the expected pionic final~state.}
	\label{fig:nnbar_general}
\end{figure}

A diagrammatic representation of an experiment involving free neutron oscillations is presented in Figure~\ref{fig:nnbar_general}. A stream of moderated (slow/cold) neutrons is directed from the source, guided through a magnetically shielded region with a length $L$\footnote{The~propagation length of neutrons would be around $200\,$m for NNBAR and $50\,$m for HIBEAM.}, and targeted at a thin carbon disk (approximately \SI{100}{\mu m} thick). Virtually 100\% of antineutrons are expected to undergo annihilation within the target.
Another method for investigating the possibility of $n \rightarrow \bar{n}$ transitions entails a different approach. In this scenario, the transformation from neutrons to antineutrons ($n \rightarrow \bar{n}$) would occur through sterile neutron transitions, as proposed in Ref.~\cite{Berezhiani:2020vbe}. Here, magnetic field scans of the volumes through which the neutrons are traveling are required, as opposed to the neutrons being in a quasi-free state, as would be needed in the experimental arrangement depicted in Figure~\ref{fig:nnbar_general}. The crucial aspect for detecting both types of searches is that, the resulting interaction of an antineutron with a nucleon in the target disk will lead to an annihilation process, which gives rise to a distinct final state involving multiple pions. This striking final state must be reconstructed by a detector surrounding it.

\subsection{The Annihilation Detector}
\begin{figure}{
  \centering
  \subfigure[]{\includegraphics[width=0.515\textwidth]{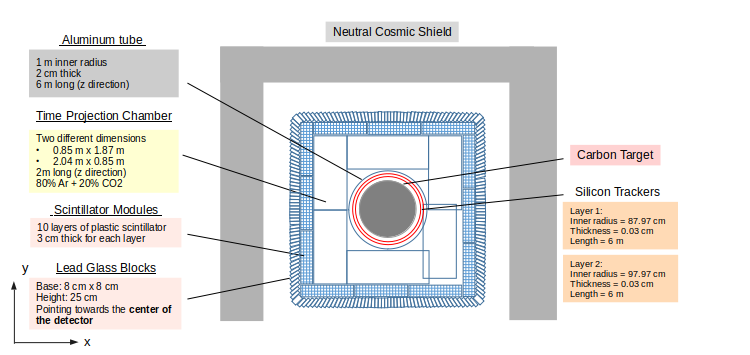}}
  \subfigure[]{\includegraphics[width=0.275\textwidth]{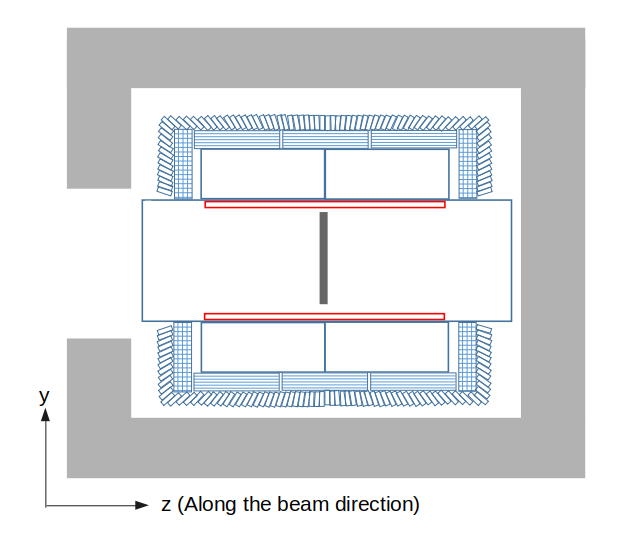}}
        \vspace*{-2mm}  %
	\caption{Schematic overview of the NNBAR detector~design in the x-y (a) and y-z (b) views.}
	\label{box_full_det_sketch}
}
\end{figure}

\vspace{-0.10cm}
Various technologies and designs are being evaluated for the annihilation detector ~\cite{Barrow:2021deh, Yiu:2022faw,Backman:2022szk}, which are being carefully investigated using the $\textsc{Geant4}$ simulation software~\cite{Yiu:2022faw, GEANT4:2002zbu}. The central components of the current baseline detector, arranged radially outward, are as follows: a) the annihilation target, a \SI{100}{\mu m} thick carbon disk, initially with a \SI{1}{\m} diameter for the HIBEAM stage and later expanded to a \SI{2}{\m} diameter for the complete NNBAR second stage; b) a charged particle tracker designed for tracking, pion identification, and determination of the annihilation vertex. The inner tracking is achieved using a silicon layer situated within a \SI{2}{\cm} thick aluminum beampipe. The beampipe is encompassed by a Time Projection Chamber (TPC), which enables particle identification by measurement of the specific continuous energy loss, $\frac{dE}{dx}$; c) for measuring charged pions, there is a hadronic range detector constructed with 10 scintillator slats arranged orthogonally. Additionally, to measure photons (essential for neutral pion reconstruction), an electromagnetic calorimeter made of lead-glass modules is employed; d) encircling the main detector, a cosmic ray background veto system based on scintillator technology will be positioned. Work is currently in progress to build a prototype detector system consisting of these components~\cite{Dunne:2021arq}, with the exception of the active cosmic veto system. 
Figure ~\ref{box_full_det_sketch} provides an overview of the NNBAR detector. It extends over \SI{600}{\cm} along the longitudinal ($z$) axis. Across the transverse ($x-y$) plane, the detector spans \SI{515}{\cm} in both width and height. The diagram includes labels denoting the various detector components and their corresponding dimensions.

\section{Conclusion}
\vspace{-0.20cm}
The HIBEAM-NNBAR experimental program for the European spallation source will conduct a series of precise experiments to search for neutron conversions using free neutrons. An improvement in sensitivity by three orders of magnitude compared to the last search is expected. These conversions, if observed, would be unmistakable evidence of baryon number violation, which addresses important questions in fundamental physics, including the baryon asymmetry of the universe. 

\bibliographystyle{elsarticle-num-names}
\bibliography{fpcp}

\begin{thebibliography}{31}
\expandafter\ifx\csname natexlab\endcsname\relax\def\natexlab#1{#1}\fi
\providecommand{\url}[1]{\texttt{#1}}
\providecommand{\href}[2]{#2}
\providecommand{\path}[1]{#1}
\providecommand{\DOIprefix}{doi:}
\providecommand{\ArXivprefix}{arXiv:}
\providecommand{\URLprefix}{URL: }
\providecommand{\Pubmedprefix}{pmid:}
\providecommand{\doi}[1]{\href{http://dx.doi.org/#1}{\path{#1}}}
\providecommand{\Pubmed}[1]{\href{pmid:#1}{\path{#1}}}
\providecommand{\bibinfo}[2]{#2}
\ifx\xfnm\relax \def\xfnm[#1]{\unskip,\space#1}\fi
\bibitem[{Addazi et~al.(2021)}]{Addazi:2020nlz}
\bibinfo{author}{A.~Addazi}, et~al.,
\newblock \bibinfo{title}{{New high-sensitivity searches for neutrons
  converting into antineutrons and/or sterile neutrons at the HIBEAM/NNBAR
  experiment at the European Spallation Source}},
\newblock \bibinfo{journal}{J. Phys. G} \bibinfo{volume}{48}
  (\bibinfo{year}{2021}) \bibinfo{pages}{070501}.
  \DOIprefix\doi{10.1088/1361-6471/abf429}.
  \href{http://arxiv.org/abs/2006.04907}{{\tt arXiv:2006.04907}}.
\bibitem[{Abele et~al.(2023)}]{Abele:2022iml}
\bibinfo{author}{H.~Abele}, et~al.,
\newblock \bibinfo{title}{{Particle Physics at the European Spallation
  Source}},
\newblock \bibinfo{journal}{Phys. Rept.} \bibinfo{volume}{1023}
  (\bibinfo{year}{2023}) \bibinfo{pages}{1--84}.
  \DOIprefix\doi{10.1016/j.physrep.2023.06.001}.
  \href{http://arxiv.org/abs/2211.10396}{{\tt arXiv:2211.10396}}.
\bibitem[{Sakharov(1967)}]{Sakharov:1967dj}
\bibinfo{author}{A.~D. Sakharov},
\newblock \bibinfo{title}{{Violation of CP Invariance, C asymmetry, and baryon
  asymmetry of the universe}},
\newblock \bibinfo{journal}{Pisma Zh. Eksp. Teor. Fiz.} \bibinfo{volume}{5}
  (\bibinfo{year}{1967}) \bibinfo{pages}{32--35}.
  \DOIprefix\doi{10.1070/PU1991v034n05ABEH002497}.
\bibitem[{Weyl(1929)}]{Weyl:1929fm}
\bibinfo{author}{H.~Weyl},
\newblock \bibinfo{title}{{Electron and Gravitation. 1. (In German)}},
\newblock \bibinfo{journal}{Z. Phys.} \bibinfo{volume}{56}
  (\bibinfo{year}{1929}) \bibinfo{pages}{330--352}.
  \DOIprefix\doi{10.1007/BF01339504}.
\bibitem[{Davoudiasl and Mohapatra(2012)}]{Davoudiasl:2012uw}
\bibinfo{author}{H.~Davoudiasl}, \bibinfo{author}{R.~N. Mohapatra},
\newblock \bibinfo{title}{{On Relating the Genesis of Cosmic Baryons and Dark
  Matter}},
\newblock \bibinfo{journal}{New J. Phys.} \bibinfo{volume}{14}
  (\bibinfo{year}{2012}) \bibinfo{pages}{095011}.
  \DOIprefix\doi{10.1088/1367-2630/14/9/095011}.
  \href{http://arxiv.org/abs/1203.1247}{{\tt arXiv:1203.1247}}.
\bibitem[{Gardner(2022)}]{Gardner:2022yeq}
\bibinfo{author}{S.~Gardner},
\newblock \bibinfo{title}{{New Opportunities for the Study of Baryon Number
  Violation at Low-Energy Accelerators}},
\newblock \bibinfo{journal}{J. Phys. Conf. Ser.} \bibinfo{volume}{2391}
  (\bibinfo{year}{2022}) \bibinfo{pages}{012016}.
  \DOIprefix\doi{10.1088/1742-6596/2391/1/012016}.
\bibitem[{Workman et~al.(2022)}]{ParticleDataGroup:2022pth}
\bibinfo{author}{R.~L. Workman}, et~al. (\bibinfo{collaboration}{Particle Data
  Group}),
\newblock \bibinfo{title}{{Review of Particle Physics}},
\newblock \bibinfo{journal}{PTEP} \bibinfo{volume}{2022} (\bibinfo{year}{2022})
  \bibinfo{pages}{083C01}. \DOIprefix\doi{10.1093/ptep/ptac097}.
\bibitem[{Davidson(1979)}]{Davidson:1978pm}
\bibinfo{author}{A.~Davidson},
\newblock \bibinfo{title}{{$B-L$ as the fourth color within an
  $\mathrm{SU}(2)_L \times \mathrm{U}(1)_R \times \mathrm{U}(1)$ model}},
\newblock \bibinfo{journal}{Phys. Rev. D} \bibinfo{volume}{20}
  (\bibinfo{year}{1979}) \bibinfo{pages}{776}.
  \DOIprefix\doi{10.1103/PhysRevD.20.776}.
\bibitem[{Rao and Shrock(1984)}]{Rao:1983sd}
\bibinfo{author}{S.~Rao}, \bibinfo{author}{R.~E. Shrock},
\newblock \bibinfo{title}{{Six Fermion ($B-L$) Violating Operators of Arbitrary
  Generational Structure}},
\newblock \bibinfo{journal}{Nucl. Phys. B} \bibinfo{volume}{232}
  (\bibinfo{year}{1984}) \bibinfo{pages}{143--179}.
  \DOIprefix\doi{10.1016/0550-3213(84)90365-1}.
\bibitem[{Mohapatra and Marshak(1980)}]{Mohapatra:1980qe}
\bibinfo{author}{R.~N. Mohapatra}, \bibinfo{author}{R.~E. Marshak},
\newblock \bibinfo{title}{{Local B-L Symmetry of Electroweak Interactions,
  Majorana Neutrinos and Neutron Oscillations}},
\newblock \bibinfo{journal}{Phys. Rev. Lett.} \bibinfo{volume}{44}
  (\bibinfo{year}{1980}) \bibinfo{pages}{1316--1319}.
  \DOIprefix\doi{10.1103/PhysRevLett.44.1316}, \bibinfo{note}{[Erratum:
  Phys.Rev.Lett. 44, 1643 (1980)]}.
\bibitem[{Nussinov and Shrock(2002)}]{Nussinov:2001rb}
\bibinfo{author}{S.~Nussinov}, \bibinfo{author}{R.~Shrock},
\newblock \bibinfo{title}{{N - anti-N oscillations in models with large extra
  dimensions}},
\newblock \bibinfo{journal}{Phys. Rev. Lett.} \bibinfo{volume}{88}
  (\bibinfo{year}{2002}) \bibinfo{pages}{171601}.
  \DOIprefix\doi{10.1103/PhysRevLett.88.171601}.
  \href{http://arxiv.org/abs/hep-ph/0112337}{{\tt arXiv:hep-ph/0112337}}.
\bibitem[{Arnold et~al.(2013)Arnold, Fornal, and Wise}]{Arnold:2012sd}
\bibinfo{author}{J.~M. Arnold}, \bibinfo{author}{B.~Fornal},
  \bibinfo{author}{M.~B. Wise},
\newblock \bibinfo{title}{{Simplified models with baryon number violation but
  no proton decay}},
\newblock \bibinfo{journal}{Phys. Rev. D} \bibinfo{volume}{87}
  (\bibinfo{year}{2013}) \bibinfo{pages}{075004}.
  \DOIprefix\doi{10.1103/PhysRevD.87.075004}.
  \href{http://arxiv.org/abs/1212.4556}{{\tt arXiv:1212.4556}}.
\bibitem[{Dev and Mohapatra(2015)}]{Dev:2015uca}
\bibinfo{author}{P.~S.~B. Dev}, \bibinfo{author}{R.~N. Mohapatra},
\newblock \bibinfo{title}{{TeV scale model for baryon and lepton number
  violation and resonant baryogenesis}},
\newblock \bibinfo{journal}{Phys. Rev. D} \bibinfo{volume}{92}
  (\bibinfo{year}{2015}) \bibinfo{pages}{016007}.
  \DOIprefix\doi{10.1103/PhysRevD.92.016007}.
  \href{http://arxiv.org/abs/1504.07196}{{\tt arXiv:1504.07196}}.
\bibitem[{Gardner and Yan(2019)}]{Gardner:2018azu}
\bibinfo{author}{S.~Gardner}, \bibinfo{author}{X.~Yan},
\newblock \bibinfo{title}{{Processes that break baryon number by two units and
  the Majorana nature of the neutrino}},
\newblock \bibinfo{journal}{Phys. Lett. B} \bibinfo{volume}{790}
  (\bibinfo{year}{2019}) \bibinfo{pages}{421--426}.
  \DOIprefix\doi{10.1016/j.physletb.2019.01.054}.
  \href{http://arxiv.org/abs/1808.05288}{{\tt arXiv:1808.05288}}.
\bibitem[{Heeck and Takhistov(2020)}]{Heeck:2019kgr}
\bibinfo{author}{J.~Heeck}, \bibinfo{author}{V.~Takhistov},
\newblock \bibinfo{title}{{Inclusive Nucleon Decay Searches as a Frontier of
  Baryon Number Violation}},
\newblock \bibinfo{journal}{Phys. Rev. D} \bibinfo{volume}{101}
  (\bibinfo{year}{2020}) \bibinfo{pages}{015005}.
  \DOIprefix\doi{10.1103/PhysRevD.101.015005}.
  \href{http://arxiv.org/abs/1910.07647}{{\tt arXiv:1910.07647}}.
\bibitem[{Girmohanta and Shrock(2020)}]{Girmohanta:2019fsx}
\bibinfo{author}{S.~Girmohanta}, \bibinfo{author}{R.~Shrock},
\newblock \bibinfo{title}{{Baryon-Number-Violating Nucleon and Dinucleon Decays
  in a Model with Large Extra Dimensions}},
\newblock \bibinfo{journal}{Phys. Rev. D} \bibinfo{volume}{101}
  (\bibinfo{year}{2020}) \bibinfo{pages}{015017}.
  \DOIprefix\doi{10.1103/PhysRevD.101.015017}.
  \href{http://arxiv.org/abs/1911.05102}{{\tt arXiv:1911.05102}}.
\bibitem[{Girmohanta(2021)}]{Girmohanta:2020eav}
\bibinfo{author}{S.~Girmohanta},
\newblock \bibinfo{title}{{Nucleon and dinucleon decays to leptonic final
  states in a left-right symmetric model with large extra dimensions}},
\newblock \bibinfo{journal}{Eur. Phys. J. C} \bibinfo{volume}{81}
  (\bibinfo{year}{2021}) \bibinfo{pages}{143}.
  \DOIprefix\doi{10.1140/epjc/s10052-021-08936-w}.
  \href{http://arxiv.org/abs/2005.12952}{{\tt arXiv:2005.12952}}.
\bibitem[{Berryman et~al.(2022)Berryman, Gardner, and
  Zakeri}]{Berryman:2022zic}
\bibinfo{author}{J.~M. Berryman}, \bibinfo{author}{S.~Gardner},
  \bibinfo{author}{M.~Zakeri},
\newblock \bibinfo{title}{{Neutron Stars with Baryon Number Violation, Probing
  Dark Sectors}},
\newblock \bibinfo{journal}{Symmetry} \bibinfo{volume}{14}
  (\bibinfo{year}{2022}) \bibinfo{pages}{518}.
  \DOIprefix\doi{10.3390/sym14030518}.
  \href{http://arxiv.org/abs/2201.02637}{{\tt arXiv:2201.02637}}.
\bibitem[{Peggs(2013)}]{Peggs:2013sgv}
\bibinfo{author}{S.~Peggs},
\newblock \bibinfo{title}{{ESS Technical Design Report}}
  (\bibinfo{year}{2013}). \URLprefix
  \url{https://europeanspallationsource.se/documentation/tdr.pdf}.
\bibitem[{Baldo-Ceolin et~al.(1994)}]{Baldo-Ceolin:1989vpk}
\bibinfo{author}{M.~Baldo-Ceolin}, et~al.,
\newblock \bibinfo{title}{{A New experimental limit on neutron - anti-neutron
  oscillations}},
\newblock \bibinfo{journal}{Z. Phys. C} \bibinfo{volume}{63}
  (\bibinfo{year}{1994}) \bibinfo{pages}{409--416}.
  \DOIprefix\doi{10.1007/BF01580321}.
\bibitem[{Kerbikov(2019)}]{Kerbikov:2018mct}
\bibinfo{author}{B.~O. Kerbikov},
\newblock \bibinfo{title}{{The effect of collisions with the wall on
  neutron-antineutron transitions}},
\newblock \bibinfo{journal}{Phys. Lett. B} \bibinfo{volume}{795}
  (\bibinfo{year}{2019}) \bibinfo{pages}{362--365}.
  \DOIprefix\doi{10.1016/j.physletb.2019.06.041}.
  \href{http://arxiv.org/abs/1810.02153}{{\tt arXiv:1810.02153}}.
\bibitem[{Nesvizhevsky et~al.(2019{\natexlab{a}})Nesvizhevsky, Gudkov,
  Protasov, Snow, and Voronin}]{Nes:2019PRL}
\bibinfo{author}{V.~V. Nesvizhevsky}, \bibinfo{author}{V.~Gudkov},
  \bibinfo{author}{K.~V. Protasov}, \bibinfo{author}{W.~M. Snow},
  \bibinfo{author}{A.~Y. Voronin},
\newblock \bibinfo{title}{{Experimental approach to search for free
  neutron-antineutron oscillations based on coherent neutron and antineutron
  mirror reflection}},
\newblock \bibinfo{journal}{Phys. Rev. Lett.} \bibinfo{volume}{122}
  (\bibinfo{year}{2019}{\natexlab{a}}) \bibinfo{pages}{221802}.
\bibitem[{Nesvizhevsky et~al.(2019{\natexlab{b}})Nesvizhevsky, Gudkov,
  Protasov, Snow, and Voronin}]{Nes:2019PLB}
\bibinfo{author}{V.~V. Nesvizhevsky}, \bibinfo{author}{V.~Gudkov},
  \bibinfo{author}{K.~V. Protasov}, \bibinfo{author}{W.~M. Snow},
  \bibinfo{author}{A.~Y. Voronin},
\newblock \bibinfo{title}{{Comment on B.O. Kerbikov "The effect of collisions
  with the wall on neutron-antineutron transitions" Phys. Lett. B 795 (2019)
  362}},
\newblock \bibinfo{journal}{Phys. Lett. B} \bibinfo{volume}{803}
  (\bibinfo{year}{2019}{\natexlab{b}}) \bibinfo{pages}{135357}.
\bibitem[{Protasov et~al.(2020)Protasov, Gudkov, Kupryianova, Nesvizhevsky,
  Snow, and Voronin}]{Pro:2020PRD}
\bibinfo{author}{K.~V. Protasov}, \bibinfo{author}{V.~Gudkov},
  \bibinfo{author}{E.~A. Kupryianova}, \bibinfo{author}{V.~V. Nesvizhevsky},
  \bibinfo{author}{W.~M. Snow}, \bibinfo{author}{A.~Y. Voronin},
\newblock \bibinfo{title}{{Theoretical analysis of antineutron-nucleus data
  needed for antineutron mirrors in neutron-antineutron oscillation
  experiments}},
\newblock \bibinfo{journal}{Phys. Rev. D} \bibinfo{volume}{102}
  (\bibinfo{year}{2020}) \bibinfo{pages}{075025}.
\bibitem[{Gudkov et~al.(2021)}]{Gudkov:2021wvn}
\bibinfo{author}{V.~Gudkov}, et~al.,
\newblock \bibinfo{title}{{A Possible Neutron-Antineutron Oscillation
  Experiment at PF1B at the Institut Laue Langevin}},
\newblock \bibinfo{journal}{Symmetry} \bibinfo{volume}{13}
  (\bibinfo{year}{2021}) \bibinfo{pages}{2314}.
  \DOIprefix\doi{10.3390/sym13122314}.
\bibitem[{Berezhiani(2021)}]{Berezhiani:2020vbe}
\bibinfo{author}{Z.~Berezhiani},
\newblock \bibinfo{title}{{A possible shortcut for
  neutron\textendash{}antineutron oscillation through mirror world}},
\newblock \bibinfo{journal}{Eur. Phys. J. C} \bibinfo{volume}{81}
  (\bibinfo{year}{2021}) \bibinfo{pages}{33}.
  \DOIprefix\doi{10.1140/epjc/s10052-020-08824-9}.
  \href{http://arxiv.org/abs/2002.05609}{{\tt arXiv:2002.05609}}.
\bibitem[{Barrow et~al.(2021)}]{Barrow:2021deh}
\bibinfo{author}{J.~Barrow}, et~al.,
\newblock \bibinfo{title}{{Computing and Detector Simulation Framework for the
  HIBEAM/NNBAR Experimental Program at the ESS}},
\newblock \bibinfo{journal}{EPJ Web Conf.} \bibinfo{volume}{251}
  (\bibinfo{year}{2021}) \bibinfo{pages}{02062}.
  \DOIprefix\doi{10.1051/epjconf/202125102062}.
  \href{http://arxiv.org/abs/2106.15898}{{\tt arXiv:2106.15898}}.
\bibitem[{Yiu et~al.(2022)}]{Yiu:2022faw}
\bibinfo{author}{S.-C. Yiu}, et~al.,
\newblock \bibinfo{title}{{Status of the Design of an Annihilation Detector to
  Observe Neutron-Antineutron Conversions at the European Spallation Source}},
\newblock \bibinfo{journal}{Symmetry} \bibinfo{volume}{14}
  (\bibinfo{year}{2022}) \bibinfo{pages}{76}.
  \DOIprefix\doi{10.3390/sym14010076}.
\bibitem[{Backman et~al.(2022)}]{Backman:2022szk}
\bibinfo{author}{F.~Backman}, et~al.,
\newblock \bibinfo{title}{{The development of the NNBAR experiment}},
\newblock \bibinfo{journal}{JINST} \bibinfo{volume}{17} (\bibinfo{year}{2022})
  \bibinfo{pages}{P10046}. \DOIprefix\doi{10.1088/1748-0221/17/10/P10046}.
  \href{http://arxiv.org/abs/2209.09011}{{\tt arXiv:2209.09011}}.
\bibitem[{Agostinelli et~al.(2003)}]{GEANT4:2002zbu}
\bibinfo{author}{S.~Agostinelli}, et~al. (\bibinfo{collaboration}{GEANT4}),
\newblock \bibinfo{title}{{GEANT4--a simulation toolkit}},
\newblock \bibinfo{journal}{Nucl. Instrum. Meth. A} \bibinfo{volume}{506}
  (\bibinfo{year}{2003}) \bibinfo{pages}{250--303}.
  \DOIprefix\doi{10.1016/S0168-9002(03)01368-8}.
\bibitem[{Dunne et~al.(2022)Dunne, Meirose, Milstead, Oskarsson, Santoro,
  Silverstein, and Yiu}]{Dunne:2021arq}
\bibinfo{author}{K.~Dunne}, \bibinfo{author}{B.~Meirose},
  \bibinfo{author}{D.~Milstead}, \bibinfo{author}{A.~Oskarsson},
  \bibinfo{author}{V.~Santoro}, \bibinfo{author}{S.~Silverstein},
  \bibinfo{author}{S.-C. Yiu},
\newblock \bibinfo{title}{{The HIBEAM/NNBAR Calorimeter Prototype}},
\newblock \bibinfo{journal}{J. Phys. Conf. Ser.} \bibinfo{volume}{2374}
  (\bibinfo{year}{2022}) \bibinfo{pages}{012014}.
  \DOIprefix\doi{10.1088/1742-6596/2374/1/012014}.
  \href{http://arxiv.org/abs/2107.02147}{{\tt arXiv:2107.02147}}.

\end{thebibliography}

\end{document}